\documentclass[aps,showpacs,amsmath,pre]{revtex4}
\usepackage{graphicx}
\usepackage{float}
\newcommand{\y }{\'{\i}}
\newcommand{\be }{\begin{equation}}
\newcommand{\ee }{\end{equation}}

\begin{document}


\title{Effects of inert species in the gas phase in a model for the catalytic oxidation of CO}
\author{G.~M.~Buend\y a}
\affiliation{Physics Department, Universidad Sim\'on Bol\y var,\\
Apartado 89000, Caracas 1080, Venezuela}
\author{P.~A.~Rikvold}
\affiliation{ Department of Physics, \\
Florida State University, Tallahassee, FL 32306-4350, USA}

\date{\today}

\begin{abstract}
We study by kinetic Monte Carlo simulations the catalytic oxidation of carbon monoxide on a surface in the presence of contaminants in the gas phase. The process is simulated by a Ziff-Gulari-Barshad (ZGB) model that has been modified to include the effect of the contaminants and to eliminate an unphysical oxygen poisoned phase at very low CO partial pressures. The impurities can adsorb and desorb on the surface, but otherwise remain inert. We find that, if the impurities cannot desorb, no matter how small their proportion in the gas mixture, the reactive window and discontinuous transition to a CO poisoned phase at high CO pressures that characterize the original ZGB model disappear. The coverages become continuous, and once the surface has reached a steady state there is no production of CO$_2$. This  is quite different from the behavior of systems in which the surface presents a fixed percentage of impurities.  When the contaminants are allowed to desorb, the reactive phas
 e appears again for CO pressures below a value 
that depends on the proportion of contaminants in the gas and on  
their desorption rate.

\end{abstract}


\maketitle
\vspace{2.5truecm}
\section{Introduction}
\label{sec:I}

Catalytic reactions on surfaces do not only have immense technological applications~\cite{bond87, chri94}, but they are also of great research interest as a medium to study a rich and complex variety of nonequilibrium phenomena. These include chaotic behavior, critical phenomena, bistability, out-of-equilibrium phase transitions, etc~\cite{imbi95, marro99}.
The catalytic oxidation of carbon monoxide (CO) on a transition metal surface has been one of the most studied reactions, because of its crucial role in industrial applications and the rich and complex behavior it exhibits. A very fruitful model to study this reaction was proposed by Ziff, Gulari, and Barshad (ZGB)~\cite{ziff86}. The deceptively simple ZGB model has been useful in describing some of the main features of the reaction that occurs on a catalytic surface, between CO and oxygen (O) to produce carbon dioxide (CO$_2$),
 in terms of a single parameter. This parameter, $y$, which represents the
probability that the next molecule arriving at the surface is a
CO, is proportional to the partial pressure of CO in the feed gas. The ZGB model  possesses a reaction window delimited by an O poisoned phase (at low $y$) and a CO poisoned phase (at higher $y$). The transition between the O poisoned and the reactive phase is continuous, and the one between the reactive  and the CO poisoned phase has been characterized as discontinuous.

Despite its success in describing important aspects of the oxidation of CO on transition metals, the ZGB model presents two fundamental differences from the experimental system. The low-$y$ continuous transition to an O poisoned phase has not been experimentally observed~\cite{ehsa89, kris92}, and under some conditions the discontinuous transition to the CO poisoned phase disappears, and the CO$_2$ production decreases continuously~\cite{ehsa89b} with increasing $y$. These aspects can be addressed without significantly sacrificing the intrinsic simplicity of the ZGB model. For example, the continuous transition at low $y$ can be eliminated by the inclusion of an Eley-Rideal-type mechanism that allows a reaction between CO molecules in the gas phase and adsorbed O atoms on the surface~\cite{meak90, tamb94} or between adsorbed O and weakly coadsorbed CO atoms at the same site~\cite{camp80, mukh09, buen09}. As will be explained later, we here address this problem by slightly cha
 nging the adsorption
  mechanism of the O molecule on the surface. Several authors have shown that the discontinuous transition can become continuous when a nonzero CO desorption rate is included~\cite{kauk89, bros92, alba92,mach05b}, or when the surface contains more than a critical percentage of inert sites~\cite{hoen00,vale00}.

Since the catalytic oxidation of CO plays a crucial role in industrial applications, any realistic approach to this reaction  must take into account the presence of impurities. Processes performed under industrial conditions greatly differ from their laboratory counterparts, usually performed in ultrahigh vacuum. Contaminants in the gas reservoir constitute an unavoidable part of most industrial processes, and generally have the undesirable effect to reduce the catalytic activity. As an example, we can mention the presence of hydrocarbons produced by the incomplete combustion of gasoline in car engines. However, most of the theoretical studies are focused on the effect of fixed impurities on the catalytic surface. In this work we use a more direct approach by letting the impurities be part of the gas mixture. The impurities can be adsorbed on  or desorbed from the surface, but otherwise they remain inert. Other authors have applied a similar approach. 
They found that the reaction window of the model shrinks with increasing contaminant concentration, and that the high-$y$ transition changes from 
discrete to continuous~\cite{bust00, schm01, hua03}. Although some of these studies introduce the possibility that the impurities desorb from the surface, they do not analyze in detail the effects of the desorption. Since different types of impurities can have different desorption rates, we consider it important to perform a study that includes in a more detailed way the effect of the desorption rates.  

The rest of this paper is organized as follows. In Section \ref{sec-2} we describe in detail how we modify the ZGB model to eliminate the unphysical continuous phase transition at low $y$, and to include the presence of impurities in the gas phase. In Section \ref{sec-3} we present our numerical results for the modified model. We choose to present the results in two separate ways. First we study a system in which the impurities remain on the surface after being adsorbed, and next we analyze separately the case that they can also desorb. In Section \ref{sec-4} we present our conclusions.

\section{Model and Simulations}
\label{sec-2}
 In this work we study the catalytic oxidation of CO on a surface, when the gas phase consists of a mixture of CO, O$_2$, and impurities X, in different proportions. The impurities can be adsorbed on the surface, and once there they remain inert. We study separately the cases in which there is no X desorption and in which there is. 
Besides including impurities in the gas phase, our model differs from the original ZGB model in the fact that the O$_2$ molecule is adsorbed on two next-nearest-neighbor (nnn) vacant sites 
(separated by $\sqrt 2$ lattice sites) 
instead of two nearest-neighbor (nn) sites~\cite{ojed11}. This simple modification brings the model closer to physical reality. On one hand it eliminates the unphysical O poisoned phase present in the original ZGB model \cite{ziff86}, and on the other hand it takes into account experimental results that suggest that oxygen atoms tend to move apart when they hit the surface in a phenomenon called hot-atom adsorption~\cite{wint96, alba94}. The elimination of the O poisoned phase can be explained as follows. In the absence of impurities, at low values of $y$ the surface is mostly covered by oxygen. The few isolated empty sites can only be filled with CO that will react with one of its nn O and leave two nn empty sites. In the original ZGB model these empty nn sites would, with 
very high probability, be filled by oxygen, eventually poisoning the surface. In our model they can be filled only with CO that will continue reacting with their neighbors. The presence of impurities does not affect this mechanism, since they also adsorb at single sites. Actually, it is possible that they can also play a part in impeding the formation of an O poisoned-surface. 
The model  is simulated on a square lattice of
linear size $L$ that represents the catalytic surface. The Monte Carlo
simulation generates a sequence of trials: CO, X, or O$_2$ adsorption or X desorption. A site $i$ is selected at 
random. If it is occupied by a X, we attempt desorption with probability $k_x$. If the site is empty, we
attempt adsorption: CO with probability $y$,  X with probability $y_x$, or  O$_2$ with probability $1-y-y_x$. These probabilities are the relative impingement rates of the molecules and are proportional to their partial pressures. 
The adsorption of CO and X each require only a single vacant lattice site. In 
contrast, adsorption of an O$_2$ molecule requires the existence of a pair of vacant nnn sites. A nnn of site $i$ is selected at random; if it is occupied the trial ends, if not the adsorption proceeds and the O$_2$ molecule is adsorbed and dissociates into two O atoms. After a CO or O$_2$ adsorption is realized, all nn pairs are checked in random order. Pairs consisting of nn CO and O atoms react: a CO$_2$ molecule is released, and the two nn sites are vacated. 
A schematic representation of this
algorithm is given by the equations,
\begin{eqnarray}
\label{eq-1}
\text{CO(g)} + \text{S} & \rightarrow & \text{CO(a)}
\nonumber \\
\text{O}_2 + 2\text{S} & \rightarrow & 2\text{O(a)}
\nonumber \\
\text{CO(a)} + \text{O(a)} & \rightarrow & \text{CO}_2\text{(g)} + 2\text{S}
\\
\text{X(g)} + \text{S} & \rightarrow & \text{X(a)}
 \nonumber \\
\text{X(a)} &\rightarrow& \text{X(g)}+ \text{S}
\nonumber 
\;
\nonumber
\end{eqnarray}
Here S represents an empty site on the surface, $\rm g$ means gas phase, and $\rm a$ means adsorbed.
The first three steps correspond to a Langmuir-Hinshelwood mechanism. 

Our simulations are performed on a square lattice of $L\times L$ sites, with $L=120$, assuming periodic boundary conditions. The
time unit is one Monte Carlo Step per Site, MCSS, during which each 
of the $L^2$ 
sites is visited once on average. Averages are subsequently taken over $10^5$
MCSS after $8\times10^4$ MCSS that are used to achieve a stationary state. 

We define coverage as the fraction of sites on the surface occupied by a 
species. We calculate the CO, O, and X coverages, and the rate of production of CO$_2$. Starting from an empty lattice, we start to take measurements once the system has reached a steady state at the pressures $y$ and $y_x$.

\section{Results} 
\label{sec-3}
We first study the case in which the impurities do not desorb. Then we study the effects of desorption.

\subsection{ No desorption of impurities, $k_x=0$}
\begin{figure}
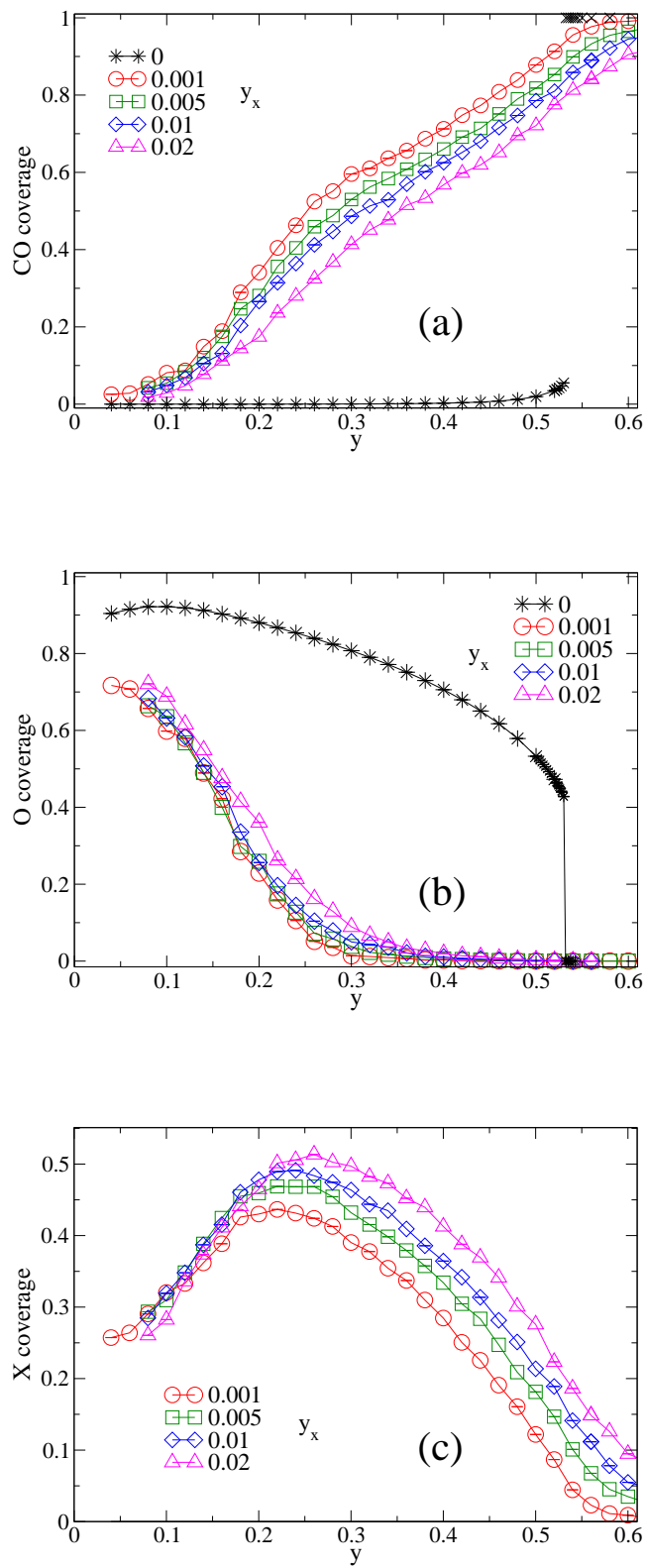

\includegraphics[scale=.35]{a120F.eps}\\
\vspace{1.5truecm}
\includegraphics[scale=.35]{b120F.eps}\\
\vspace{1.5truecm}
\includegraphics[scale=.35]{c120F.eps}\\
\caption[]{(Color on line) Coverages vs $y$ for the case that there is no impurity desorption, $k_x=0$, for the values of $y_x$ indicated in the figures. (a) CO coverage, (b) O coverage, (c) X coverage.  The reaction rate is not plotted because is always zero, except in the case that there are no impurities, i.e., $y_x=0$.} 
\label{f1}
\end{figure}

\begin{figure}
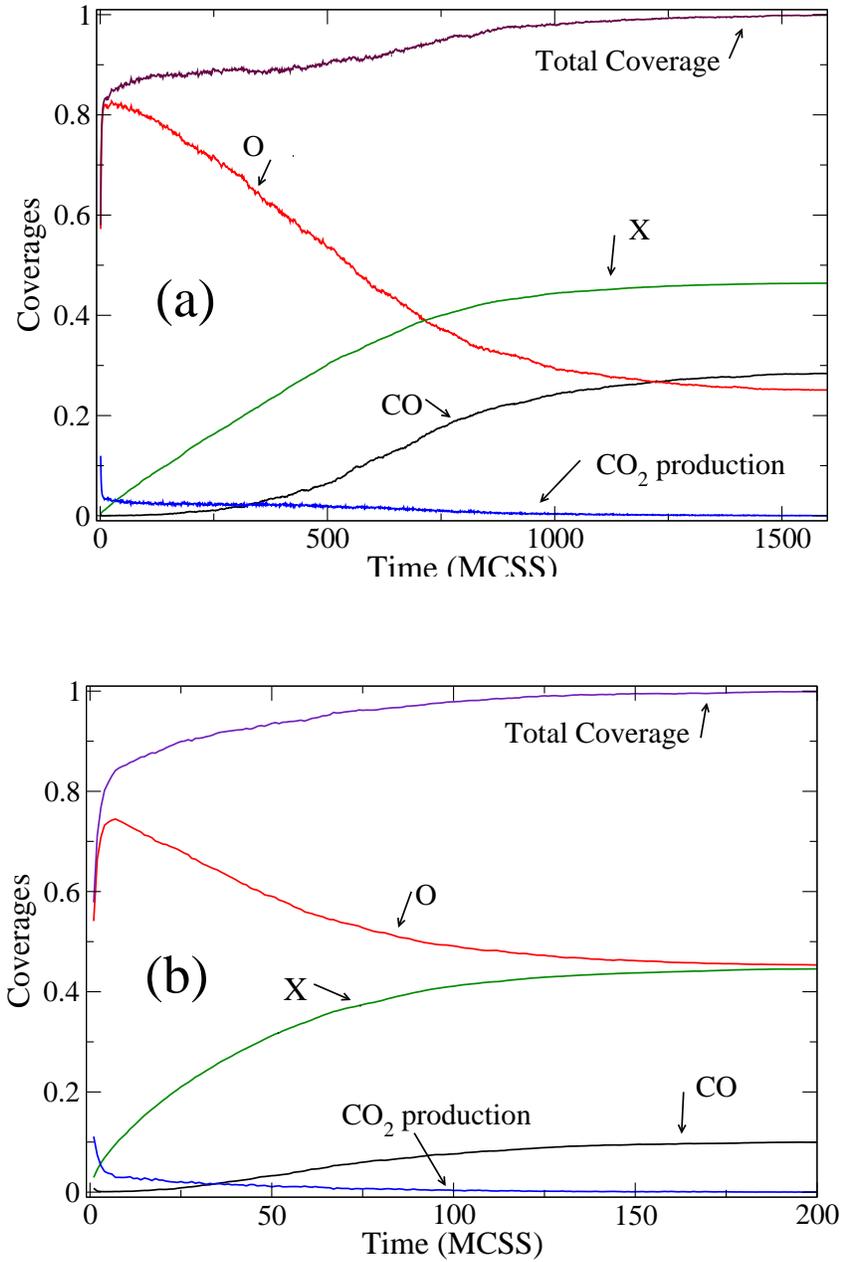

\centering\includegraphics[scale=.45]{pyx005.eps}\\
\vspace{1.2 truecm}
 \centering\includegraphics[scale=.45]{pyx05.eps}\\
\caption[]{(Color on line) CO, O, and X coverages, and CO$_2$ production rate as functions of time (measured in MCSS, starting from an empty lattice) 
for $y=0.2$, $k_x=0$, when (a) $y_x=0.005$, and (b) $y_x=0.05$. In both cases, when the system reaches the steady state the total coverage is unity (there are no empty sites on the surface), and the CO$_2$ production rate is zero. 
The production rate decays to zero after only a few MCSS.
Note the different time scales in parts (a) and (b). The system with the higher value of $y_x$ reaches the steady state much faster.}
\label{f2}
\end{figure}

\begin{figure}
\centering\includegraphics[scale=.55]{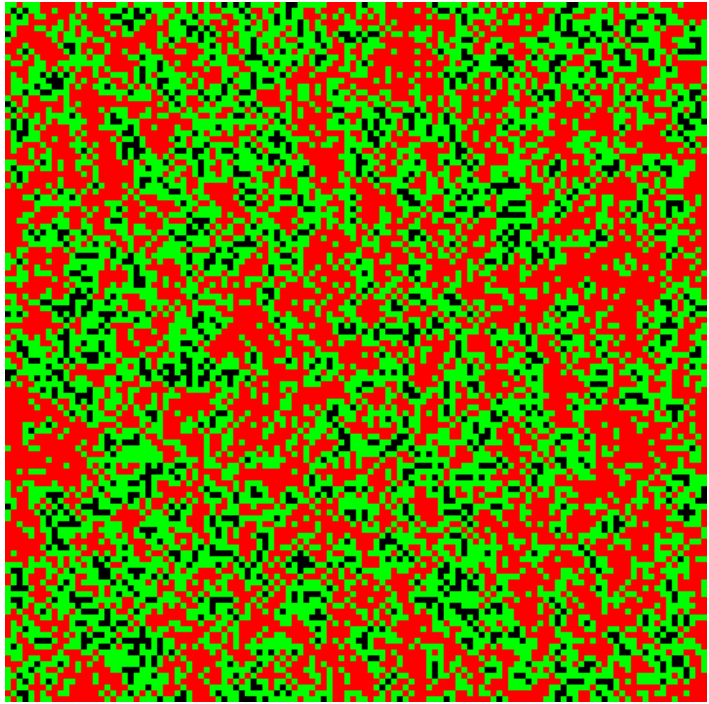}
\caption[]{(Color online) Snapshot of the surface once the stationary state has been reached. Sites filled with CO are black, with O are dark Gray (red online), and with impurities are light gray (green online), and empty sites would be white. Notice that there are no empty sites. $k_x=0$, $y=0.2$, and $y_x=0.05$ as in Fig.~\protect\ref{f2}(b).}
\label{f3}
\end{figure}

In this subsection we study the case in which the impurities remain stuck on the surface once they have been adsorbed. Consequently, the last step in the reaction described by Eq.~(\ref{eq-1}) does not occur. The effect of this type of impurities in the phase diagram of the system can be seem in Fig.~\ref{f1}, which shows the coverage of CO (a), O (b), and  X (c) vs $y$, for several values of the impurity concentration $y_x$. Notice that the continuous transition at low $y$ characteristic of the original ZGB model is absent in our model (even in the case that there are no impurities, $y_x=0$). As it was explained in the previous section, this is due to the modification in the adsorption mechanism of O. 
In a previous work it was shown that, except for the disappearance of the 
continuous phase transition, the new model behaves in a similar way as the 
original ZGB model \cite{ojed11}. The only significant quantitative difference is that that the discontinuous transition to CO poisoning occurs at a slightly 
higher value of $y$, $y_2\approx 0.530$ compared with $y_2\approx 0.526$ 
for the original ZGB model \cite{bros92}.  
The diagrams show that as soon as impurities are added to the gas mixture, the discontinuous transition to CO poisoning observed in the original ZGB model disappears: the CO and O coverages change smoothly between their extreme values. The X coverage as a function of $y$ reaches a maximum that increases with $y_x$, and then decreases. As $y_x$ increases, the location of the maximum shifts slightly toward higher values of $y$. We did not plot the reaction rate because it is different from zero only when there are no impurities, $y_x=0$. Even a minimal concentration of impurities in the gas has the effect to completely eliminate the production of CO$_2$ in the steady state. This is in clear contrast with the effect of fixed inert impurities (or inert sites) located at random on the surface. In the latter case it was found that there is a critical concentration of impurities, above which the transition becomes continuous, and that the production rate decreases continuously  
with the concentration of inert sites~\cite{hoen00, lore02}. The reason for this markedly different behavior is quite obvious. Having the relative concentration of impurities in the gas phase fixed is not equivalent to fixing the percentage of impurities on the lattice. In our model impurities continuously arrive to the surface. The system reaches a steady state in which there are no empty sites on the surface, and therefore no reactions are possible. This effect can be seen in Fig.~\ref{f2}, where we plot the time dependence of the coverages, the production rate, and the total surface coverage, for a relatively low partial pressure of impurities, Fig.~\ref{f2}(a), and a higher one, Fig.~\ref{f2}(b). In both cases it is clear that there is nonzero CO$_2$ production only during the transient part of the process. 
The CO production decays to zero after just a few MCSS. 
The steady state is reached when all the surface sites are filled (with CO, X, or O) in such a way that there are no nn pairs of type CO and O. A snapshot of the surface once it has reached the steady state is shown in Fig.~\ref{f3}. Notice how the impurities act as a barrier between the CO and O sites, preventing them from reacting. It has been shown that, when CO and O are allowed to diffuse on the surface, there can be a moderate reaction rate, even in the case that the impurities do not desorb~\cite{schm01}.

\subsection{Non-zero impurity desorption, $k_x > 0$}
\begin{figure}
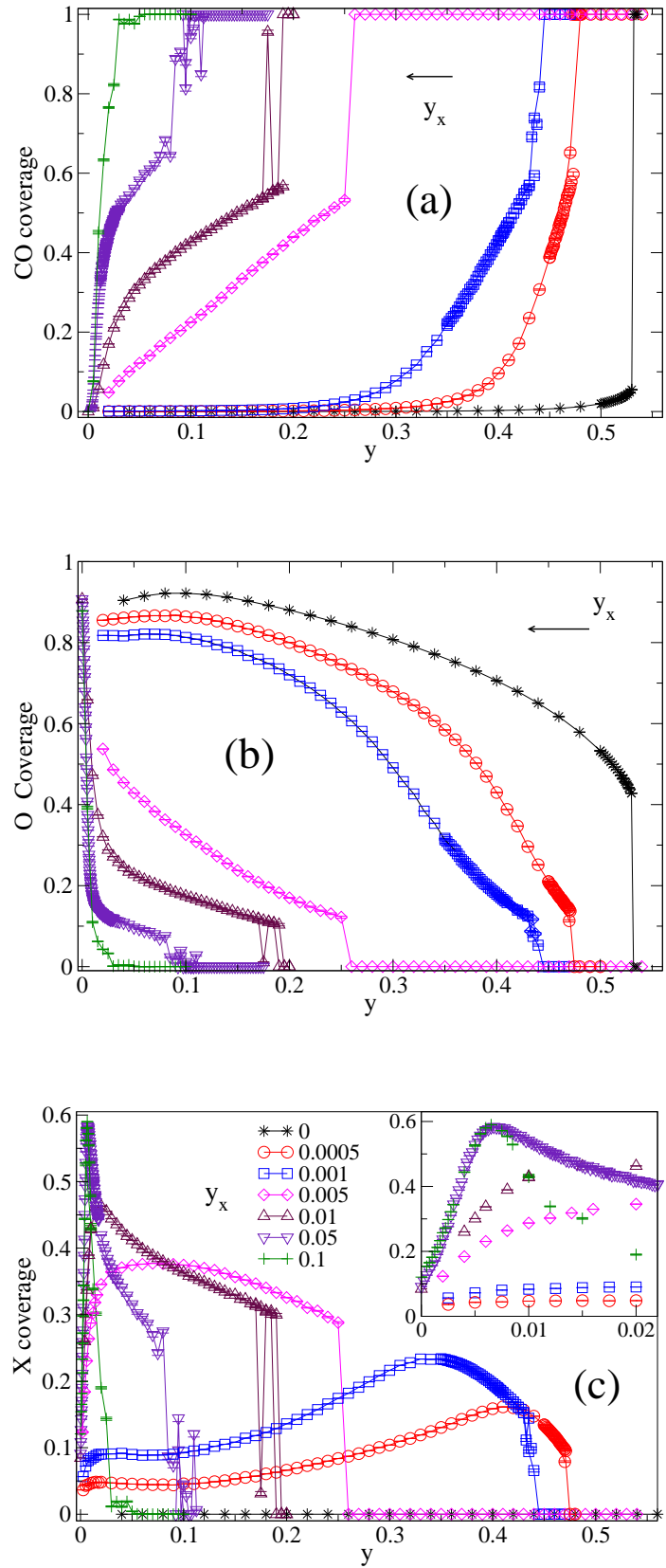

\centering\includegraphics[scale=.38]{kx001aF.eps}\\
\vspace{1.2truecm}
\centering\includegraphics[scale=.38]{kx001bF.eps}\\
\vspace{1.2truecm}
\centering\includegraphics[scale=.38]{kx001cF.eps}\\
\vspace{1.2truecm}
\caption[]{(Color online) Coverages vs $y$ when $k_x=0.001$ for different values of $y_x$. (a)CO coverage (b) O coverage (c) X coverage. The values of $y_x$ are indicated in part (c). The inset in part (c) shows a magnification  for very small $y$.}
\label{f4}
\end{figure}

\begin{figure}
\centering\includegraphics[scale=.45]{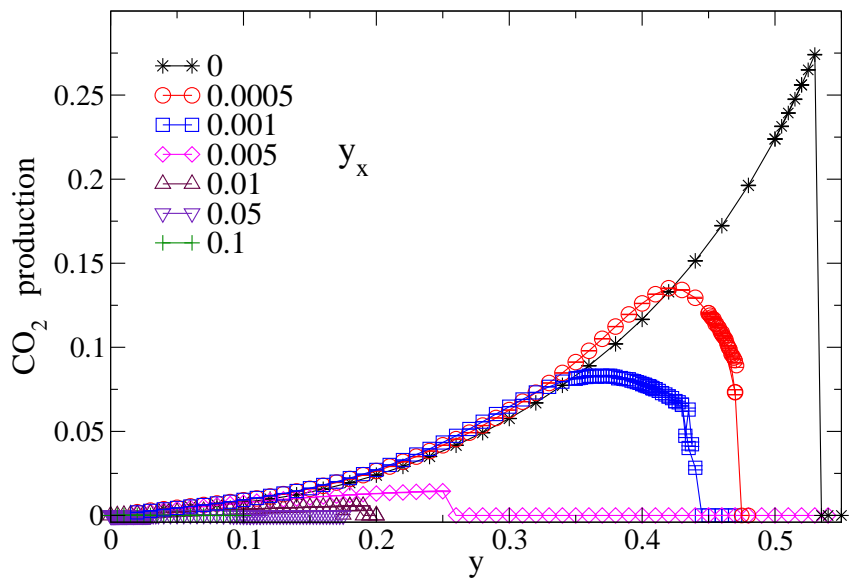}
\caption[]{(Color online) Production rate of CO$_2$ vs $y$ for several values of $y_x$ at $k_x=0.001$.}
\vspace{1.2truecm}
\label{f5}
\end{figure}

In this section we study the changes in the phase diagram when the last step of Eq.~(1) is included. Now the impurities can be desorbed from the surface with probability $k_x > 0$.  We start by fixing the desorption rate and changing the relative pressure of the impurities, $y_x$. Figures \ref{f4} and \ref{f5} show the coverages and reaction rates, respectively, for a system with $k_x=0.001$ for several values of $y_x$. Figure \ref{f4}(a) shows the CO coverage, Fig.~\ref{f4}(b) the O coverage, and Fig.~\ref{f4}(c) the X coverage. As before, 
the unphysical O poisoned phase is absent. In  Fig.~\ref{f5} we show the reaction rate. The crucial difference from the previous case of $k_x=0$ is that now the discontinuous transition reappears, and the system has a reactive phase. From the figures it is evident that as $y_x$ increases, the phase transition is shifted toward lower values of $y$, such that the reaction window is greatly reduced. The point at which the discontinuous transition from a reactive to a CO
  poisoned non-reactive phase occurs, $y_c$, now depends on both $k_x$ and $y_x$. Furthermore, for large values of $y_x$ (at fixed $k_x$) the discontinuous transition seems to disappear, and the coverages change continuously between their extreme values. This critical value of $y_x$ depends on $k_x$; from the figures it seems that $y_x^c(k_x=0.001)\approx 0.05$. In Fig.~\ref{f5} we see an interesting effect. For small enough values of $y_x$ and $y$, the reaction rates seem almost independent of $y_x$. Actually, close to the transition a very small proportion of impurities seems to {\it increase\/} the reaction rate very slightly. But as $y_x$ increases, the CO$_2$ production dramatically drops to insignificant levels. 
Notice that, independently of the concentration of impurities in the gas 
phase, $y_x$, at sufficiently large $y$
the system always reaches a state where the surface is 
completely covered with CO. 
The presence of impurities simply facilitates the emergence of this state. 
After 
 an impurity is desorbed, the empty site will more probably be filled with a CO than with an O, since the latter requires the existence of another empty site 
in an nnn position relative to the first. In Fig.~\ref{f4}(c) we see that 
the amount of impurities on the surface reaches a maximum value that, 
as expected, increases with increasing $y_x$. The location of the maximum 
moves toward lower values of $y$ as $y_x$ increases. As $y_x$ approaches 
its critical value $y_x^c(k_x)$ (above which the discontinuous 
transition disappears) 
the maximum is located very close to $y=0$ (see inset in Fig.~\ref{f4}(c)).

\begin{figure}
\centering\includegraphics[scale=.4]{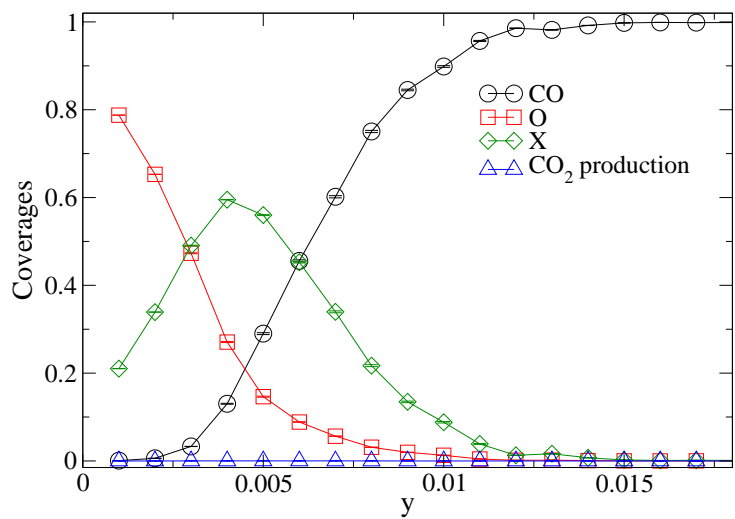}
\caption[]{(Color online) Coverages vs $y$ for the case of $y_x=0.15 > y_c(k_x)$, $k_x=0.001$. Notice that there is no discontinuous transition, and that the reaction rate is basically zero for this large value of $y_x$. This behavior has no analog in the original ZGB model.}
\label{f6}
\end{figure}

In Fig.~\ref{f6} we show the coverages vs $y$ for a case in which the discontinuous transition has disappeared. The system goes continuously to a phase in which the surface is completely filled with CO at a very small value of $y$. 
The CO$_2$ production rate in this case is negligible. 

\begin{figure}
\centering\includegraphics[scale=.4]{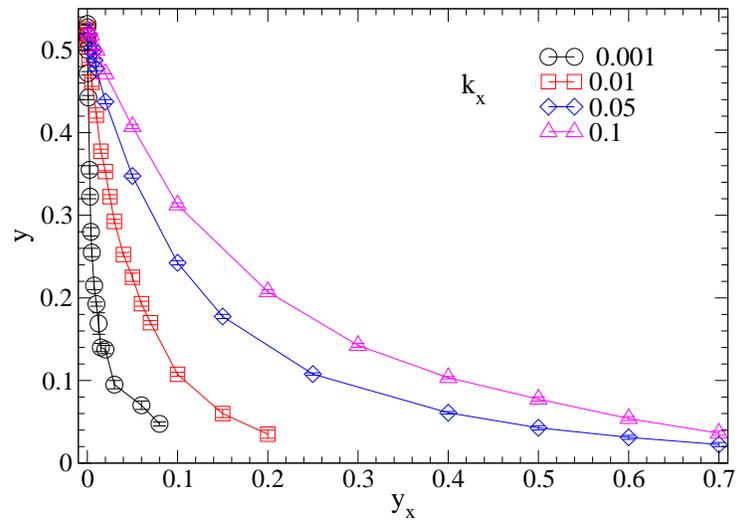}
\caption[]{(Color online) Phase diagram showing the critical value of $y$, 
below which the discontinuous transition disappears, $y_{\rm c}$ vs $y_x$,  for several values of $k_x$. $y_{\rm c}$ depends on $y_x$ and $k_x$. For $y > y_c$ the reaction rate is zero.}
\label{f7}
\end{figure}

In Fig.~\ref{f7} we plot a phase diagram showing the value of $y$, below which the discontinuous transition occurs, $y_c$, as a function of $y_x$ for several values of $k_x$. For a fixed value of $k_x$, the value of $y_c$  decreases in a non-linear way with increasing $y_x$. For a particular value of $y_x$ the transition point increases as the value of $k_x$ increases. 
The curves end at the critical value, $y_x^c(k_x)$. 
For values of $y_x$ beyond this point, the discontinuous transition 
is replaced by a smooth crossover.

\section{Conclusions}
\label{sec-4}
In this work we explore the effects of the presence of impurities in the gas phase on the reaction CO+O $\rightarrow$ CO$_2$ on a catalytic surface. Our model is a modified version of the ZGB model, in which the oxygen atoms adsorb at two nnn sites instead of two nn sites. This modification has the effect to eliminate the unphysical oxygen poisoned phase.  The impurities can be adsorbed on the surface, and once there they do not react. We found that if the impurities do not desorb, the stationary reactive phase disappears. Even the smallest percentage of impurities in the gas phase has the effect of stopping the CO$_2$ production. When a desorption rate for the impurities is included, the discontinuous transition to the CO poisoned state shifts toward lower values of $y$, such that the reaction window of the system is greatly reduced. The discontinuous transition seems to disappear at sufficiently large values of $y_x$. 
The limiting value depends on the impurity desorption rate, $k_x$. 
It is important to point out that the eventual disappearance of the reaction rate is not due to the surface being filled with impurities, but because by filling and liberating single empty sites, the impurities alter the topology of the adsorb ate layer, favoring the nucleation and growth of CO islands that eventually fill the entire surface. Compared with the original ZGB model, this model with impurities in the gas phase and with a different entrance mechanism for the oxygen molecule, is of greater experimental relevance because the unphysical oxygen-phase is absent. Also, the phase diagram is richer as, depending on the values of $y_x$ and $k_x$, there are regions where there is no CO production, and regions where the discontinuous transition to a CO poisoned phase disappears completely. We believe that this study can be useful to understand the behavior of heterogeneous catalytic CO oxidation under different industrial conditions, where the mixed gas phase can contain different types of impurities that can have different desorption rates.

\section*{Acknowledgments}
G.M.B is grateful for the hospitality of the Physics Department at Florida State University. P.A.R acknowledges support by U.S. National Science Foundation Grant Nos. DMR-0802288 and DMR-1104829.

\vspace{2.truecm}

\end{document}